\documentstyle[sprocl,epsfig]{article}

\input{psfig}

\def\Journal#1#2#3#4{{#1} {\bf #2}, #3 (#4)}

\def\be{\begin{equation}}
\def\ee{\end{equation}}
\def\bea{\begin{eqnarray}}
\def\eea{\end{eqnarray}}

\begin{document}

\title{ALTERNATIVE SCENARIOS FOR THE FRAGMENTATION OF A GLUONIC LUND STRING}

\author{BO ANDERSSON, FREDRIK S\"ODERBERG, SANDIPAN MOHANTY\footnote{talk presented by Sandipan Mohanty}}

\address{Department of Theoretical Physics, Lund University,\\
S\"olvegatan 14 A, 223 62 Lund,Sweden\\E-mail: sandipan@thep.lu.se}


\maketitle\abstracts{
The assumptions in the Lund model suffice to
prescribe a unique stochastic process for the fragmentation of
 a string into a set of hadrons, so long as the string is "flat",
 ie as long as the state described by the string consists only
 of a quark and an antiquark stretching a constant
 force field between them. Emission of gluons causes the string
 to trace more complicated surfaces in Minkowski space, and some
 form of generalization of the 1+1 dimensional model is required.
 One such generalizaiotn has been developed and implemented as a
 Monte Carlo routine "JETSET" by Torbj\"orn Sj\"ostrand, which
 has been hightly successful in describing experimental data.
 But there are theoretical reasons to believe that the
 fragmentation scheme employed in JETSET is not entirely
 satisfactory; most notably, non-adherance to the Lund Area law,
 and certain problems in handling transverse momenta. A few
 alternative scenarios, which we have examined in detail and
 implemented in separate computer programs, will be presented here,
 with comparisons to JETSET in certain simple cases. Our effort has
 been to preserve the area law for the fragmentation of a gluonic
 string, while we explored the possibility of allowing the
 fragmentation process to reshape the string surface slightly.
}

\section{Introduction}

The Lund model uses the decay of a massless relativistic string as a model for
 hadronization. Breaking of a string into a set of hadrons is thought of as a
 stochastic process with a constraint that the hadrons must be on the mass shell,
 a condition of saturation and an assumption of symmetry between the
 quark and the antiquark ends of the string. These assumptions lead \cite{bgb,lm} to a unique
 solution for the probabilities involved in string breaking. The probability for
a certain hadronic state to emerge by the fragmentation of a string turns out
to be proportional to the hadronic phase space and the negative exponential of
the area spanned by the string in space-time before it decays.

\section{Gluonic strings}
\subsection{Directrix}
	 The principle of least action applied to a string gives a
minimal surface as the world surface of the string, described by a boundary
curve called directrix. The assumption that the string doesn't have any
longitudinal degrees of freedom, and that the string tension has
the same value in the rest frame of a local piece of string, lead to the
conclusion that the directrix is a curve which has a light like tangent every
where.

	The directrix is determined in terms of the initial conditions on the string.
Normally we are interested in a string state that was produced from a few
quarks and gluons originating at a point or from a very small volume where
perturbative effects dominate. So, we can think of the initial conditions on
the string to be a few colour connected lumps of energy with different energy
momentum vectors, which were created at the same point in space at some time.
This initial condition leads to a very simple directrix : Partonic energy
momenta arranged one after another in colour order.

\subsection{Fragmentation as a process along the directrix} Since the string is
completely described in terms of the directrix, it should be possible to describe
all its properties, including how it fragments, as processes along the directrix.
Without going into the details, let's just note that the following algorithm
implements fragmentation of a string as a process along the directrix and is
completely equivalent, for gluonless strings, to the iterative process used for
instance in JETSET.

Let's define:
$x_n=\sum_{i=1}^{n} p_i$,
where $p_i$ is the momentum of the i'th rank particle, and an associated
vector $q_n$ for the n'th breakup vertex. (for the gluonless case, this is
just the vector position of the breakup point).

\begin{itemize}
	\item Initialize $q_0 = k_q$, and x=0.
	\item pick a random number z from the distribution
$f(z)=\frac{(1-z)^a}{z}e^{-\frac{bm^2}{z}}$.
	\item starting from the tip of the vector $(x_i+q_i)$, which is a point on
the directrix, find a segment k along the directrix, such that
$kq_i=\frac{m^2}{z}$, and k is lightlike.
	\item find particle momentum p, and next q from :
$p_{i+1}= \frac{1}{2} l+\frac{1}{2} k$, and
$q_{i+1}=q_i- \frac{1}{2} l +\frac{1}{2} k$
 where,
$l= 2z(q_i-\frac{q_i^2}{2q_i k} k)$
	\item update $x_{i+1}=x_i+p_i$, and the point on the directrix marked by x+q
	\item repeat all steps (except initialization!)
\end{itemize}
\newpage
\begin{center}
\begin{figure}[h]
\begin{center}
\psfig{figure=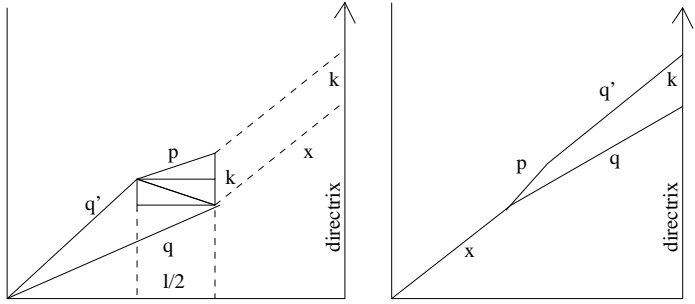, height=4cm}
\caption{A fragmentation step between vertices at q and q'. Particle momentum p is the sum of
a fraction z of the light-like vector $q-\frac{q^2}{2qk} k$ (written as $\frac{l}{2}$
in the text) and $\frac{k}{2}$. The figure on right shows the same step with the vector
x drawn from the origin instead of the vectors $q_i$. }
\end{center}
\end{figure}
\end{center}

        This algorithm implements area law in the following way.. The area of
the quadilaterlal formed by $q_i,k,q_{i+1}$ and $p$, an area associated with
production of one particle, is the same as the area of the flat triangular region
between $q_i$ and $q_{i+1}$ (translated to the origin) plus a half the hadron
mass. So the area between
the curve traced by the particle momenta arranged in a chain and the directrix is
equal to the area spanned by the string before it decays. Building up the area
traced by the string in this way, was used in making analytical calculations in the
Lund model\cite{bf}.

The above expressions for $q_i$, $q_{i+1}$ and p are completely symmetric with
respect to forward and backward steps along the directrix. One can define
$\bar{z}$ such that,

$k=2\bar{z}(q_{i+1}-\frac{q_{i+1}^2}{2q_{i+1} l} l)$, and $lq_{i+1} =
\frac{m^2}{\bar{z}}$.

The relation between z and $\bar{z}$ can be found from the above:
$\bar{z}=\frac{\frac{m^2}{z}}{q_1^2 +\frac{m^2}{z}}$, which inverts to,
$z=\frac{\frac{m^2}{\bar{z}}}{q_2^2 +\frac{m^2}{\bar{z}}}$.

The square of the vectors $q_i$ evolve as,
$q_{i+1}=(1-z)(q^2+\frac{m^2}{z})$

Eventually the vector $x+q$ will reach close to a gluon cornor, a place where the
directrix changes direction. The remaining
straight segment (which we will call 'c'), might be such that
$cq<\frac{m^2}{z}$. It is then impossible to select a k that is a
fraction of c and fullfills $kq=\frac{m^2}{z}$.

\section{Fragmentation of a gluonic string}

	For a string with gluons, one can't assume that any region of the string is
equivalent to any other region except for scaling and Lorentz transformations.
The energy momenta of the decay products are not simple linear combinations of
two light like vectors. The assumptions mentioned at the beginning of this
article, therefore, do not lead to any definite scheme for fragmentation of a
gluonic string.
\subsection{Method used in JETSET}
	To solve the problem the following additional inputs were made into the
Monte Carlo program JETSET :

\begin{itemize}
	\item The string state as determined by perturbation theory (e.g. the dipole
 cascade model), is considered to be frozen during the fragmentation process. That is,
 the process of hadronization doesn't interfere with the structure of the string.

	\item z is to be reinterpreted as a connection between $q_{i+1}^2$ and
$q_i^2$ through the relation $q_{i+1}=(1-z)(q_i^2+\frac{m^2}{z})$, where $q_i$s
are the vector positions of the fragmentation vertices, on the string surface.
\end{itemize}

This last relation and $p^2=m^2=-(\Delta q)^2$ always uniquely determine a
point $q_{i+1}$ on the string surface. This means that the vertices obtained for the
flat string by area law fragmentation, are projected on to the surface of the multigluon
state. Unfortunately the area below the vertices is not preserved by this projection.
That is, the probability for decay of the string into a cluster of
hadrons is no longer proportional to the area spanned by the string before it
decays.

\subsection{Alternatives: Modifying the directrix}

	We start by recalling that a string state is a record of the quarks and
gluons present in a hadronization environment. In particular, we note that
large flat areas on the string surface correspond to large squared mass
between two partons, and small areas correspond to low mass between two
partons. At the point where our process along the directrix stopped, we were
left with a mass $cq_i$ which was of the order of a hadron mass. It was
impossible to find a k that was along the directrix, was light-like, and had
the required product with q, because the available lightlike segment of the
directrix was too short.

	But the directrix is not calculated to infinite precision. Parton shower in
the dipole cascade model for instance, is continued down to a certain
scale in gluonic transverse momentum. The smaller this scale, the more gluons
we would have. And at about the energy scale of hadron
masses, a new physical phenomenon takes place: hadrons form. So, the detailed structure of the
string resolved to the
hadronization scale, might perhaps be affected by the process by which hadrons form.

	In the following, this idea that the dipole cascade model determines a string
state approximately, and the small scale details are fixed dynamically by the
process of hadronization, has been used to break the deadlock in the fragmentation
algorithm at gluon cornors. We think of the segments k in each step of the
fragmentation process outlined earlier, as the segments of the real directrix.

	As long as these can be chosen along the perturbatively calculated directrix,
there is no problem. This can't be done when the directrix changes
direction, as mentioned earlier. Since k can not be chosen as a fraction of c,
we take c and another segment of the directrix, K (not necessarily lightlike),
and express their sum as two different lightlike vectors. That is, we write,
$c+K=k+c'$, such that k and c' are lightlike. We demand also that
$kq=\frac{m^2}{z}$, so that we can make one more fragmentation step using this k.

	These conditions do not determine the vectors k and c' uniquely, since the
length of the vector K is also unknown. By adding one more condition these
vectors can be fixed. We tested several choices for this condition...

\begin{itemize}
\item eg2: Choose k as the unique light-like vector in the plane spanned by q and Q
which is closer to Q than q.
\item eg3: Choose k such that c and $c'$ have the same lengths in the rest frame
of q
\end{itemize}

        We note that for both these choices, it may happen that the modification of
the directrix will contain sharp "conrors". Such situations will break the coherence
conditions in a perturbative QCD cascade, in particular, the requirement of strong
angular ordering.
\begin{itemize}
\item eg4: Same as eg3 except for the provision that 3-vector part of $c'$ would
not be allowed have an angle in excess of 90 degrees with the next local part
of the old directrix.
\end{itemize}
        It is possible to introduce further "smoothness conditions" to get rid of these problems but
we have found that the following choice solves the problem everywhere.
\begin{itemize}

\item eg5: Chose $c'$ such that after the production of a particle using k, $c'$
can be used for the production of another particle, so that after two steps
the vector x+q is back on the directrix. That is, the required additional
condition is $c'q'=\frac{m^2}{z'}$, where $q'$ is the q vector obtained after
the production step involving k.
\end{itemize}

	We summarize in figures 2, the results of using these methods and
         compare them to a curve
produced using JETSET in one simple situation: a perturbative string state
consisting of a quark, an antiquark and a single gluon. We plot the rapidity
distributions obtained from the different methods here. It is interesting to
see that the fragmentation procedure produces too many particles over a large region (cf figure 2)
although the modification of the directrix is actually only of the order of the
hadron mass scale everywhere.

\begin{figure} [h]
\psfig{figure=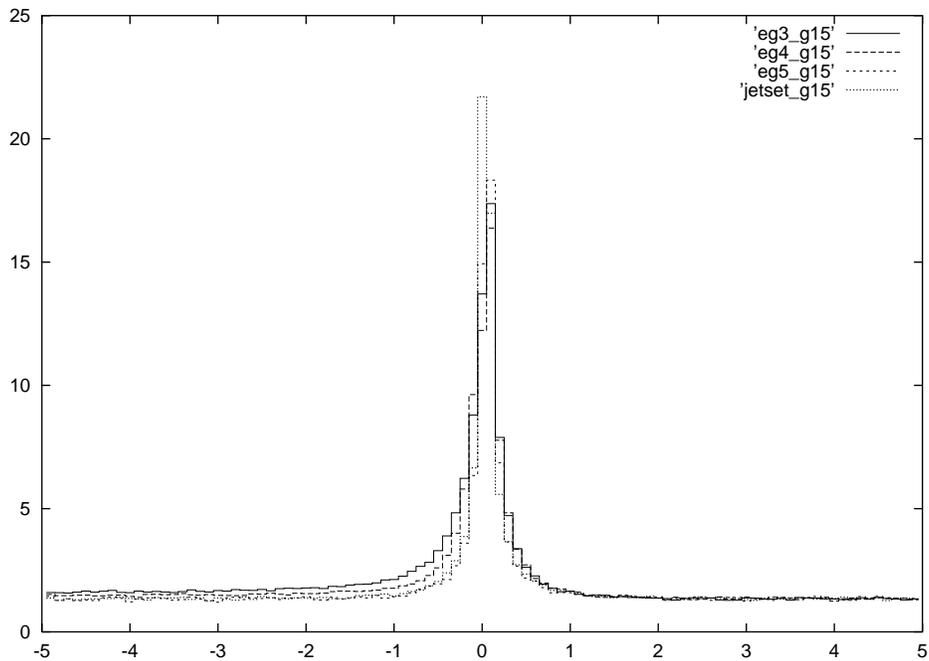,height=9cm}
\caption{Rapidity distribution for a 3 jet event obtained from
different fragmentation proceedures mentioned in the text. }
\end{figure}
\begin{figure}[h]
\psfig{figure=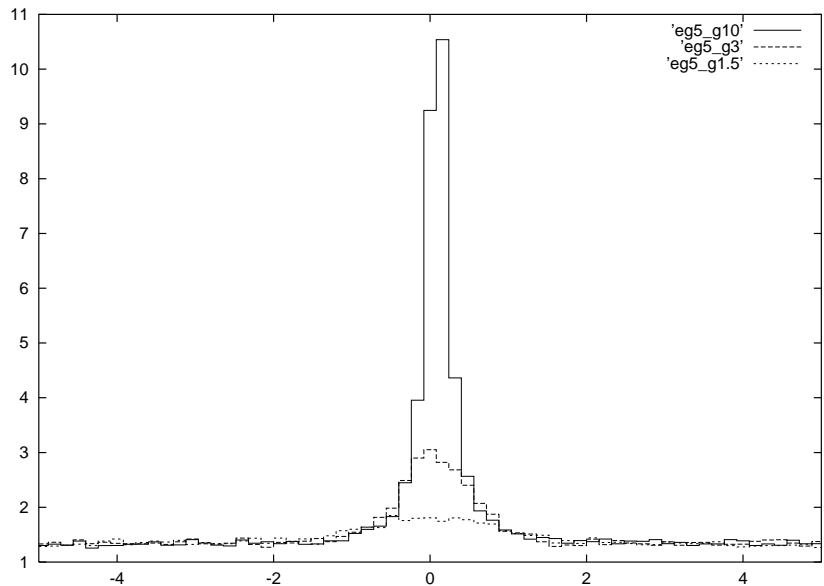,height=8cm}
\caption{Rapidity distribution for a 3 jet event from eg5, for gluon transverse momentum values 10,3 and 1.5 GeV  }
\end{figure}
\newpage
\section{Conclusions}
        We reported at the meeting that the additional soft gluons which we introduced to effectuate area law fragmentation
have to have an angular ordering property. We have since the time of the meeting learned many more things (to be discussed
in a forthcoming paper). But our major finding is that any modification of the directrix must be such that the bremstahlung
coherence conditions of perturbative QCD must be fulfilled even if the modifications are chosen to be small and local. The
method indicated at the end of the talk could be chosen as a general strategy for these modifications.

\end{document}